\documentstyle[12pt]{article}
\begin{document}
\newcommand{\vsp}{\vspace*{3mm}}
\newcommand{\be}{\begin{equation}}
\newcommand{\ee}{\end{equation}}
\newcommand{\bd}{\begin{displaymath}}
\newcommand{\ed}{\end{displaymath}}
\newcommand{\bra}{\langle}
\newcommand{\ket}{\rangle}
\newcommand{\order}{{\cal O}}
\newcommand{\plus}{\!+\!}
\newcommand{\minus}{\!-\!}
\title{\bf Closure of Macroscopic Laws\\ in Disordered Spin Systems:\\
a Toy Model}
\date{}
\author{ A.C.C. Coolen \dag \and  S. Franz \ddag}
\maketitle
\begin{center}
\dag Department of Physics - Theoretical Physics, University of Oxford\\
1 Keble Road, Oxford OX1 3NP, U.K.
\end{center}
\begin{center}
\ddag Nordita\\
Blegdamsvej 17, DK-2100, Copenhagen, Denmark
\end{center}
\vsp

\begin{center}{PACS: 75.10N, 87.30, 5.20}\end{center}
\vsp

\begin{abstract}
We use a linear system of Langevin spins with disordered
interactions as an exactly solvable toy model to investigate
a procedure, recently  proposed by Coolen and Sherrington, for closing the
hierarchy of
macroscopic order parameter equations in disordered spin systems.
The closure procedure, based on the removal of microscopic
memory effects, is shown to reproduce the correct
equations for short times and in equilibrium. For intermediate
time-scales the procedure does not lead to the exact equations, yet for
homogeneous initial conditions succeeds at capturing the main
characteristics of the flow in the order parameter plane.
The procedure fails in terms
of the long-term temporal dependence of the order parameters.
For low energy inhomogeneous initial conditions and near criticality (where
zero
modes appear) deviations in temporal behaviour are most apparent.
For homogeneous initial conditions the impact of microscopic memory effects
on the evolution of macroscopic order parameters in disordered spin
systems appears to be mainly an overall slowing down.
\end{abstract}

\pagebreak
The off-equilibrium dynamics of mean field
 spin-glass models has been a central subject of research in recent times
\cite{cuku1,frme1,frme2,cuku2}.
Starting from a stochastic Markovian dynamics for the microscopic variables,
it is found that the  appropriate "order parameter functions" - the
spin-spin correlation function at different times, and the associated
response function - obey non Markovian equations. These contain  at
all times
"memory terms" which depend on the previous history.  It has been recently
emphasized that these memory terms play an important role at low temperatures,
being responsible for asymptotic breaking of time translation invariance and
aging.

If one wants to retain a description of the dynamics at a Markovian level,
it is possible to write down  an exact
hierarchy of equations which does not close.
An infinite
number of instantaneous "order parameters" is therefore required.
Recently Coolen and Sherrington \cite{CS,CSSK} (CS) proposed a method for
obtaining a
closed set of autonomous  macroscopic differential equations, by a systematic
elimination of microscopic memory effects. The crucial simplifying hypothesis
is that at any time step the system is at equilibrium on the surface
where some macroscopic variables related to the energy and the
magetization are constant.
 They applied this method to the
Hopfield \cite{hopfield} model and the
Sherrington-Kirkpatrick (SK) \cite{SK} model, which are the archetypical models
for attractor neural
networks and spin-glasses, respectively. The CS procedure,  by
construction, gives the exact equations at $t=0$ and
the correct equilibrium fixed points. In addition it is exact for all
times in the limit where the disorder is
removed. For the SK model and the Hopfield model, however, there are
no sufficiently detailed and reliable analytical or numerical results to allow
for a direct
test of the
procedure at intermediate times, although recent data on cumulants of the
local field distribution in the Hopfield model do suggest
deviations between theory and simulations \cite{ON}.
For the Hopfield model, for instance, non-trivial finite
size effects are known to persists even in systems of size
$N\sim 10^6$ \cite{kohring}. Secondly, in both models a full
analysis following CS requires going through a dynamic version of the replica
symmetry breaking
scheme a la Parisi \cite{parisi}, for a continuous range of times and
parameters, which is in practice unattainable.

Nevertheless, it is worth noticing that even if  the CS procedure
neglects the possibility of aging - the
crucial aspect of the spin-glass dynamics -
the "flow diagrams" obtained for the order
parameters in \cite{CS,CSSK}
show qualitative agreement
with the results of numerical simulations.
In order to understand the potential and the restrictions of the
closure procedure proposed by CS, we present in this letter results of
 studying an exactly solvable toy
model, for which replica symmetry in the dynamical equations is
stable. Since the CS procedure is based on the elimination
of microscopic memory effects, this also allows us to obtain a better
understanding of the
role of these effects in determining the macroscopic
behaviour of disordered spin systems.
\vsp

As our toy model we choose a linear system of $N$ Langevin spins
$\{\sigma_i\}$ with disordered interactions $\{J_{ij}\}$ and a
Gaussian white noise $\{\eta_i(t)\}$:
\be
\frac{d}{dt}\sigma_i=\sum_{j=1}^N J_{ij}\sigma_j - \mu\sigma_i
+\eta_i~~~~~~
\bra\eta_i(t)\eta_j(t^{\prime})\ket=2T\delta_{ij}\delta(t-t^{\prime})
\label{eq:langevin}
\ee
In which the symmetric interactions $\{J_{ij}\}$
are drawn at random from a Gaussian distribution with $\bra
J_{ij}\ket=0$ and $\bra J_{ij}^2\ket=JN^{-\frac{1}{2}}$.
For large
$N$ the eigenvalue distribution $\rho(\lambda)$ of the interaction matrix is
given by
Wigner's semi-circular law \cite{wigner}
\be
\rho(\lambda)=\frac{\sqrt{4J^2-\lambda^2}}{2\pi
J^2}\theta\left[2J+\lambda\right]\theta\left[2J-\lambda\right]
\ee
so we have to choose $\mu\geq2J$ in order to suppress runaway modes.
Both the statics and the dynamics of this linear model are trivially solved.
In statics, it is found that the replica symmetric solution is always stable
and the model does not have a glassy phase. The system is always in a "high
temperature" condition. Here we can expect that the memory effects play only a
minor role, compared to spin-glass systems which exhibit a transition.
 The dynamics
is solved by transformation to the basis where the interaction
matrix is diagonal, i.e. $\sigma_i(t)\rightarrow\sigma_{\lambda}(t)$
and $\eta_i(t)\rightarrow\eta_{\lambda}(t)$ (this transformation does
not affect the statistical properties of the noise). Alternatively,
one could write coupled equations for correlation- and
response-functions,
containing memory terms.
Following the former route  one obtains
\be
\sigma_{\lambda}(t)= \sigma_{\lambda}(0)e^{-t(\mu-\lambda)}
+\int_0^t\!ds~\eta_{\lambda}(s)e^{(\mu-\lambda)(s-t)}
\ee
In particular we can calculate directly the quantitities in terms of which
the CS procedure will be formulated,  the average spin norm
$Q\equiv\frac{1}{N}\sum_i\sigma_i^2$ and the energy per spin
$E\equiv-\frac{1}{2N}\sum_{ij}\sigma_i
J_{ij}\sigma_j+\frac{\mu}{N}\sum_i\sigma_i^2$:
\be
Q(t)=\int\!d\lambda~\rho(\lambda)\sigma_{\lambda}^2(0)e^{-2t(\mu-\lambda)}+T
\int\!d\lambda\frac{\rho(\lambda)}{\mu-\lambda}\left[1-e^{-2t(\mu-\lambda)}\right]
\label{eq:Q(t)}
\ee
\be
E(t)=\frac{1}{2}\int\!d\lambda~\rho(\lambda)\sigma_{\lambda}^2(0)(\mu-\lambda)e^{-2t(\mu-\lambda)}+\frac{1}{2}T\left[1-\int\!d\lambda~\rho(\lambda)e^{-2t(\mu-\lambda)}\right]
\label{eq:E(t)}
\ee
We will (improperly)  call these  quantities "order parameters" for the system.
The macroscopic equilibrium state is found to be
\be
Q(\infty)=\frac{T}{2J^2}\left[\mu-\sqrt{\mu^2-4J^2}\right]~~~~~~E(\infty)=\frac{1}{2}T
\label{eq:equilibrium}
\ee
We will consider two types of initial conditions.
For homogeneous initial conditions, $\sigma_{\lambda}^2(0)=Q_0$, we
can write the solution (\ref{eq:Q(t)},\ref{eq:E(t)}) in the compact form
\be
Q(t)=Q_0-4\int_{0}^t\!ds\left[E(s)-\frac{1}{2}T\right]
\ee
\be
E(t)=\frac{1}{2}T-\frac{1}{2}\left[\frac{1}{2}Q_0\frac{d}{dt}+T\right]\frac{e^{-2\mu
t} I_1(4Jt)}{2Jt}
\label{eq:homogeneous}
\ee
in which $I_1(z)$ denotes the modified Bessel function
\cite{abramowitz}. We can use the properties of
$I_1(z)$ to obtain from (\ref{eq:homogeneous}) directly the short-time and
the asymptotic behaviour of the system, for comparison with the
results of the CS procedure. For short times we find
\bd
E(t)=\frac{1}{2}\mu Q_0+t\left[\mu T-Q_0(\mu^2+J^2)\right]
-t^2\left[T(\mu^2+J^2)-\mu Q_0(\mu^2+3J^2)\right]
\ed
\be
+\frac{2}{3}t^3\left[\mu T(\mu^2+3J^2)-Q_0(\mu^4+6\mu^2
J^2+2J^4)\right]+\order(t^4)~~~~~~(t\rightarrow0)
\label{eq:shortexacthomo}
\ee
whereas the asymtotic behaviour turns out to be described by
\be
E(t)=\frac{1}{2}T-\frac{e^{-2t(\mu-2J)}}{4\sqrt{\pi}(2Jt)^{3/2}}\left[T-Q_0(\mu-2J)+\order(\frac{1}{t})\right]~~~~~~(t\rightarrow\infty)
\ee
The second type of initial conditions we consider are
inhomogeneous ones, where the system is prepared in one specific
eigendirection $\Lambda$ of the interaction matrix, so
$\sigma_{\lambda}^2(0)=Q_0\delta(\lambda-\Lambda)\rho^{-1}(\Lambda)$.
The solution (\ref{eq:Q(t)},\ref{eq:E(t)}) can now be written as
\be
Q(t)=Q_0-4\int_{0}^t\!ds\left[E(s)-\frac{1}{2}T\right]
\ee
\be
E(t)=\frac{1}{2}Q_0(\mu-\Lambda)e^{-2t(\mu-\Lambda)}+
\frac{1}{2}T\left[1-\frac{e^{-2\mu
t} I_1(4Jt)}{2Jt}\right]
\label{eq:inhomogeneous}
\ee
For short times we now find
\bd
E(t)=\frac{1}{2}(\mu-\Lambda) Q_0+t\left[\mu T-Q_0(\mu-\Lambda)^2\right]
-t^2\left[T(\mu^2+J^2)-Q_0(\mu-\Lambda)^3\right]
\ed
\be
+\frac{2}{3}t^3\left[\mu
T(\mu^2+3J^2)-Q_0(\mu-\Lambda)^4\right]+\order(t^4)~~~~~~(t\rightarrow0)
\label{eq:shortexactinhomo}
\ee
whereas the asymtotic behaviour is given by
\be
E(t)=\frac{1}{2}T+\frac{1}{2}(\mu-\Lambda)Q_0 e^{-2t(\mu-\Lambda)}
-\frac{T
e^{-2t(\mu-2J)}}{4\sqrt{\pi}(2Jt)^{3/2}}\left[1+\order(\frac{1}{t})\right]~~~~~~(t\rightarrow\infty)
\ee
For $\mu>2J$ there are no zero modes and the asymptotic relaxation is
simply
exponential, with characteristic time
$\tau=\left[2(\mu-2J)\right]^{-1}$. For $\mu=2J$, however, zero modes
appear, as a result of which we find power laws:
\be
Q(t)-Q(\infty)\sim t^{-\frac{1}{2}}~~~~~~~~E(t)-E(\infty)\sim t^{-3/2}
\ee
The only asymptotic diffence between the two types of initial
conditions is that in the case of choosing a zero mode as the initial
state ($\mu=\Lambda=2J$), the equilibrium norm $Q(\infty)$ will depend
on $Q_0$.
\vsp

We now turn to the CS procedure \cite{CS,CSSK} for deriving a closed set of
deterministic macroscopic differential equations.
The total energy per spin is separated into two contributions, one
of which depends on the realisation of the disorder, and one
of which does not.
These two quantities (or equivalently any functions thereof) will evolve in
time deterministically on finite
time-scales and are
chosen to represent a macroscopic state. For the present model
we can choose $Q$ and $E$. From the Fokker-Planck equation associated
with (\ref{eq:langevin}) follows  a Liouville equation for the macroscopic
probability
distribution ${\cal P}_t(Q,E)$, which describes the deterministic flow
\be
\frac{d}{dt}Q=-4\left[E-\frac{1}{2}T\right]
\label{eq:dQdt}
\ee
\be
\frac{d}{dt}E=\mu T-\bra h^2 \ket_{Q,E;t}
\label{eq:dEdt}
\ee
with the sub-shell average
\be
\bra h^2 \ket_{Q,E;t}\equiv
\frac{\int\!d\vec{\sigma}~p_t(\vec{\sigma})~
\delta\left[Q-Q(\vec{\sigma})\right]
\delta\left[E-E(\vec{\sigma})\right]
\frac{1}{N}\sum_i\left[\sum_j
J_{ij}\sigma_j +\mu\sigma_i\right]^2}
{\int\!d\vec{\sigma}~p_t(\vec{\sigma})~
\delta\left[Q-Q(\vec{\sigma})\right]
\delta\left[E-E(\vec{\sigma})\right]}
\label{eq:localfielddist}
\ee
These laws are exact, although not yet closed due to the
appearance of the microscopic probability distribution
$p_t(\vec{\sigma})$ in (\ref{eq:localfielddist}). In the case of the
Hopfield \cite{hopfield} and the Sherrington-Kirkpatrick \cite{SK}
model, removing the disorder (by putting $\alpha=0$ and $\tilde{J}=0$
in these models, respectively) closes the hierarchy \cite{CS,CSSK}.
The same happens in the present toy model: for $J=0$
the sum over sites in (\ref{eq:localfielddist}) simply equals $\mu^2
Q$, the microscopic distribution $p_t(\vec{\sigma})$ drops out and
the equations
(\ref{eq:dQdt},\ref{eq:dEdt}) close.

Following CS we now close the equations (\ref{eq:dQdt},\ref{eq:dEdt})
for arbitrary $J$ by assuming $(i)$ selfaveraging of the flow with
respect to the microscopic realisation of the disorder, and $(ii)$
that in evaluating the disorder-averaged sub-shell average $\bra h^2
\ket_{Q,E;t}$
(\ref{eq:localfielddist}) we may assume equipartitioning of probability
within the $(Q,E)$ sub-shells of the ensemble.
As a result $\bra h^2 \ket_{Q,E;t}$ is replaced by
\be
\bra h^2 \ket_{Q,E}\equiv\bra
\frac{\int\!d\vec{\sigma}~
\delta\left[Q\minus Q(\vec{\sigma})\right]
\delta\left[E\minus
E(\vec{\sigma})\right]
\frac{1}{N}\sum_i\left[\sum_j
J_{ij}\sigma_j +\mu\sigma_i\right]^2}
{\int\!d\vec{\sigma}~
\delta\left[Q\minus Q(\vec{\sigma})\right]
\delta\left[E\minus E(\vec{\sigma})\right]}\ket_{\{J_{ij}\}}
\label{eq:equipfielddist}
\ee
The set (\ref{eq:dQdt},\ref{eq:dEdt}) is now closed and the
sub-shell average (\ref{eq:equipfielddist}) can be calculated with the replica
method. From this stage onwards all calculations can be performed
exactly. The underlying assumptions are guaranteed to be
exact at $t=0$ (upon
choosing appropriate initial conditions) and in equilibrium (as a result
of the Boltzmann form of the equilibrium distribution).

In calculating  (\ref{eq:equipfielddist}) with the
replica method there enters an
 auxiliary spin-glass type order
parameter $q(Q,E)$, with the physical meaning
\be
q\equiv\bra
\frac
{\int\!\int\!d\vec{\sigma}d\vec{\sigma}^{\prime}~\frac{1}{N}\sum_i\sigma_i\sigma_i^{\prime}~
\delta\left[Q\minus Q(\vec{\sigma})\right]
\delta\left[E\minus E(\vec{\sigma})\right]
\delta\left[Q\minus Q(\vec{\sigma}^{\prime})\right]
\delta\left[E\minus E(\vec{\sigma}^{\prime})\right]
}
{\int\!\int\!d\vec{\sigma}d\vec{\sigma}^{\prime}~
\delta\left[Q\minus Q(\vec{\sigma})\right]
\delta\left[E\minus E(\vec{\sigma})\right]
\delta\left[Q\minus Q(\vec{\sigma}^{\prime})\right]
\delta\left[E\minus E(\vec{\sigma}^{\prime})\right]
}\ket_{\{J_{ij}\}}
\label{eq:q}
\ee
Upon making the replica symmetry (RS) ansatz one finds three different regions
in the $(Q,E)$
plane, characterised by different associated values of $q$ and of the
relevant sub-shell average $\bra h^2 \ket_{Q,E}$, the boundaries of which are
the
lines $E=E_A$ and $E=E_B$:
\be
E_A\equiv\frac{1}{2}Q(\mu\minus J)
{}~~~~~~~~
E_B\equiv\frac{1}{2}Q(\mu\plus J)
\ee
By calculating the eigenvalue
$\lambda_{\rm AT}$, which determines the stability of the RS saddle
point in  the so-called replicon direction \cite{AT}, we find that RS
is truly stable in the $q=0$ region ($\lambda_{\rm AT}<0$), and
marginally stable in the two $q>0$ regions ($\lambda_{\rm AT}=0$):
\be
\begin{array}{ccccccc}
{\rm region} & & q & & \lambda_{\rm AT} & & dE/dt \\ \\
E<E_A & &  >0 &  & 0 & &
-2\mu(E\minus\frac{1}{2}T)+(\mu\minus 2J)(\mu Q\minus 2E) \\ \\
E_A < E < E_B & & 0 & & <0 & & \mu T-QJ^2-4E^2/Q \\ \\
E>E_B & & >0 & & 0 & &
-2\mu(E\minus\frac{1}{2}T)+(\mu\plus 2J)(\mu Q\minus 2E)\\
\end{array}
\label{eq:finaldEdt}
\ee
In the limit of zero disorder we indeed recover the correct (trivial)
evolution for the remaining order parameter $Q$.
We will now asses to which degree the CS flow equations
(\ref{eq:dQdt},\ref{eq:finaldEdt}) reproduce or approach the exact
results in the presence of disorder.
\vsp

According to (\ref{eq:dQdt},\ref{eq:finaldEdt}) the flow in the two $q>0$
regions is directed into the
middle region $q=0$, where the fixed-point of the flow is indeed given by the
correct expression (\ref{eq:equilibrium}). At criticality ($\mu=2J$),
however, the $q=0$ fixed-point is precisely on the regional boundary
$E=E_A$ and a degenerated stable line $E=\frac{1}{2}T$ develops in the region
$E<E_A$.
Solving numerically the flow equations
(\ref{eq:dQdt},\ref{eq:finaldEdt}) results in figure \ref{fig:CS}
(in this example $\mu=8$ and $T=3$), where we show the flow iterated
for $0\leq t \leq 10$
from initial states which are drawn either homogeneously (on the line
$E=\frac{1}{2}\mu Q_0$) or inhomogeneously from the extreme modes
$\lambda=\pm 2J$ (on the lines $E=\frac{1}{2}Q_0(\mu\pm 2J)$, which are
the boundaries of the physical region).
\begin{figure}
\vspace*{23cm}
\hbox to \hsize{\hspace*{-4cm}\includegraphics{CS.ps}\hspace*{4cm}}
\vspace*{-15cm}
\caption{Flow in the $(Q,E)$ plane according to the CS equations, for
$\mu=8$ and $T=3$. Left picture: $J/\mu=3/8$ (no zero modes), right
picture: $J/\mu=1/2$ (zero modes).  Outer two
dashed lines: boundaries of the physical region,
$E=\frac{1}{2}Q(\mu\pm 2J)$ (note: for $J/\mu=1/2$ one of these
 coincides with the line $E=0$). Inner two dashed lines: boundaries
$E_{A,B}$ of the $q=0$ region. Thin horizontal line segment in right figure:
degenerated stable line $E=\frac{1}{2}T$.}
\label{fig:CS}
\end{figure}
\begin{figure}
\vspace*{23cm}
\hbox to \hsize{\hspace*{-4cm}\includegraphics{exact.ps}\hspace*{4cm}}
\vspace*{-15cm}
\caption{Flow in the $(Q,E)$ plane according to the exact solution, for
$\mu=8$ and $T=3$. Left picture: $J/\mu=3/8$ (no zero modes), right
picture: $J/\mu=1/2$ (zero modes).  Outer
dashed lines: boundaries of the physical region,
$E=\frac{1}{2}Q(\mu\pm 2J)$ (note: for $J/\mu=1/2$ one of these
coincides with the line $E=0$). Inner dashed lines: boundaries
$E_{A,B}$ of the $q=0$ region.}
\label{fig:exact}
\end{figure}
The corresponding flow according to the exact equations
(\ref{eq:homogeneous}) and (\ref{eq:inhomogeneous}) is shown in
figure \ref{fig:exact}.
Away from the critical situation $\mu=2J$
there is a qualitative agreement between the two flows, especially for
homogeneous initial conditions. For $\mu=2J$ there are
clear deviations in the region $E<E_A$ (where one finds the zero
modes). In this picture we have added flow lines starting
from initial states with $E=\frac{1}{2}Q_0(\mu-
2J)+\epsilon$ $(\epsilon\ll1)$, to emphasize the difference with the
zero modes $E=\frac{1}{2}Q_0(\mu-
2J)$.
The  degenerated line $E=\frac{1}{2}T$ of the CS equations is
in reality not stable.

\begin{figure}
\vspace*{23cm}
\hbox to \hsize{\hspace*{-4cm}\includegraphics{homotimes.ps}\hspace*{4cm}}
\vspace*{-15cm}
\caption{Comparison of order parameter evolution starting from
homogeneous initial states, for
$\mu=8$, $T=3$ and $J/\mu=3/8$ (no zero modes). Solid lines: exact
equations, dashed lines: CS closure.}
\label{fig:homotime}
\end{figure}
\begin{figure}
\vspace*{23cm}
\hbox to \hsize{\hspace*{-4cm}\includegraphics{inhomotimes.ps}\hspace*{4cm}}
\vspace*{-15cm}
\caption{Comparison of order parameter evolution starting from
inhomogeneous initial states, for
$\mu=8$, $T=3$ and $J/\mu=3/8$ (no zero modes). Solid lines: exact
equations, dashed lines: CS closure.}
\label{fig:inhomotime}
\end{figure}
If we inspect the temporal behaviour away from the critical situation
$\mu=2J$, we find a remarkable agreement
for the case of homogeneous initial conditions (see figure
\ref{fig:homotime}). In the case of inhomogeneous
initial conditions (see figure \ref{fig:inhomotime}) there is a
difference between starting from high energy initial states $\lambda=2J$,
where there is again agreement between exact and CS
results, and starting from low energy initial states $\lambda=-2J$, where
significant deviations occur.
Expansion of the
flow equations (\ref{eq:dQdt},\ref{eq:finaldEdt}) around the
equilibrium state gives the leading asymptotic
temporal behaviour, which can be compared to the exact results (with
the dimensionless quantity $x\equiv 2J/\mu$):
\be
\begin{array}{lllll}
\mu>2J: & & E(t)\minus E(\infty)\sim e^{-t/\tau} & & Q(t)\minus
Q(\infty)\sim e^{-t/\tau} \\  \\
& & {\rm exact:} & & (2\mu\tau)^{-1}=1\minus x \\
& & {\rm CS:} & & (2\mu\tau)^{-1}=\frac{1}{2}(1\minus
x)\plus\frac{1}{2}\sqrt{1\minus x^2} \\ \\
\mu=2J: & & E(t)\minus E(\infty)\sim t^{-\alpha-1} & & Q(t)\minus Q(\infty)\sim
t^{-\alpha} \\ \\
& & {\rm exact:} & & \alpha=1/2 \\
& & {\rm CS:} & & \alpha=1 \\
\end{array}
\label{eq:asymptotics}
\ee
The agreement obtained for homogeneous
initial conditions, in spite of the difference
in characteristic relaxation times (\ref{eq:asymptotics}), can be
explained by studying the behaviour of $\log[Q-Q(\infty)]$ and
$\log[E-E(\infty)]$ as a function of time for the exact solution
(\ref{eq:homogeneous}). It turns out that the
regime of exponential relaxation sets in only extremely close to
equilibrium (for $\log[Q-Q(\infty)]$ and $\log[E-E(\infty)]$ of the
order of $10^{-15}$). For short times one can expand the CS equations
in powers of $t$ and compare with the exact expansions
(\ref{eq:shortexacthomo},\ref{eq:shortexactinhomo}), with the following
results:
\be
\begin{array}{lll}
E(0)=\frac{1}{2}\mu Q_0: &  & E_{\rm CS}(t)=E_{\rm
exact}(t)+\order(t^3) \\ \\
E(0)=\frac{1}{2}(\mu\pm 2J)Q_0: &  & E_{\rm CS}(t)=E_{\rm
exact}(t)+\order(t^2)\end{array}
\ee
which explains the difference in agreement between the two types of
initial conditions, i.e. between figures \ref{fig:homotime} and
\ref{fig:inhomotime}.
\vsp

The aim of our study was to perform a test of the closure
procedure proposed in \cite{CS,CSSK}, which is based on the removal of
any memory effects in the evolution of macroscopic quantities.
For such a test
we have chosen a model which is  exactly solvable, and does not
involve the full hierarchy of replica symmetry breaking in the dynamical
calculations of
CS.
Our study
shows that the CS procedure does not lead in any non-trivial case to the exact
equations, yet in some cases succeeds at capturing the main
characteristics of the flow in the order parameter plane.
The procedure fails in terms
of the long-term temporal dependence of the order parameters. However,
in absence of
zero modes, and
for homogeneous initial conditions the true asymptotic regime turns
out to have only restricted relevance, as it sets in extremely
late. This
implies that for homogeneous initial conditions the effect of
microscopic memory effects on the evolution of macroscopic order
parameters results  mainly in an overall slowing down.
\vsp\vsp

\noindent{\bf Aknowledgement:}\\
One of the authors, A.C., would like to thank Nordita for their
hospitality.
\vsp

\end{document}